\documentclass[conference]{IEEEtran}
\usepackage{cite}
\usepackage{amsmath,amssymb,amsfonts}
\usepackage{algorithmic}
\usepackage{graphicx}
\usepackage{textcomp}
\usepackage{xcolor}
\usepackage{overpic}
\usepackage{comment}
\usepackage{placeins}
\usepackage{pgf}
\usepackage{siunitx}

\usepackage{stackengine}
\usepackage{tikz}
\usepackage{booktabs}
\usepackage{microtype}
\usepackage{multirow}
\usepackage{makecell}  

\DeclareSIUnit{\sccm}{sccm}
\DeclareSIUnit{\ohm}{\ensuremath{\Omega}}

\newcommand{\hfald}{tetrakis(ethyl\-methyl\-amino)\allowbreak hafnium(IV)}
\newcommand{\zrald}{bis(methyl\-cyclo\-penta\-dienyl)\allowbreak(methyl)\allowbreak(methoxy)\allowbreak zirconium(IV)}

\usepackage{float}
\usepackage[hidelinks]{hyperref}
\newcommand*{\doi}[1]{\href{https://doi.org/#1}{doi:#1}}

\renewcommand{\figurename}{Fig.} 

\def\BibTeX{{\rm B\kern-.05em{\sc i\kern-.025em b}\kern-.08em
    T\kern-.1667em\lower.7ex\hbox{E}\kern-.125emX}}

\begin{document}
\bstctlcite{BSTcontrol}

\title{Back-End-Of-Line Integration of 2-Terminal Ferroelectric Hafnia Synaptic Weights on Commercial CMOS Process}

\author{
Alexander Flasby\textsuperscript{1*},
Anwesha Panda\textsuperscript{1}, 
Nathan Savoia\textsuperscript{1}, 
Laura Bégon-Lours\textsuperscript{1}\\
\textsuperscript{1}Integrated Systems Laboratory, ETH Zurich, Switzerland, 
*Corresponding author: flasbya@iis.ee.ethz.ch
}

\maketitle

\vspace{-6mm}

\begin{abstract}
Functionalising the back-end-of-line (BEOL) of CMOS with emerging materials is a key route to neural network accelerators and neuromorphic systems. In this work, 
we demonstrate for the first time the integration of 2-terminal ferroelectric synaptic weights in the BEOL of as-delivered commercial \qty{180}{\nano\metre} CMOS dies. 
A novel material stack is proposed, in which a WO\textsubscript{x}/[HfO\textsubscript{2}-ZrO\textsubscript{2}] bilayer is sandwiched between TiN and Mo electrodes. For devices annealed at \qty{450}{\celsius}, programming below \qty{1.8}{\volt} is achieved with a cycle-to-cycle variation below \qty{3}{\percent}. Synaptic weights annealed at \qty{500}{\celsius} exhibit more gradual weight updates, while the tested underlying NMOS transistors remain functional after the complete integration process. The trade-off between thermal budget and on-device training performance is illustrated using hardware-aware neural network simulations.

\end{abstract}


\section{Introduction}

Artificial Neural Networks (ANNs) are structured into layers of nodes or neurons, interconnected through matrices of synaptic weights. Training an artificial neural network is achieved by adjusting the latter. Realising these weights in hardware allows in-memory computation, circumventing the von Neumann bottleneck and accelerating the computation\cite{haensch2023compute}. One approach leverages Ohm's law and Kirchhoff's current law to execute the multiply and accumulate (MAC) operation directly in memory, requiring resistive elements that are both linear and reprogrammable \cite{sebastian2020memory}. Such ``memristors" have been successfully demonstrated in ANN accelerators, but still face integration challenges such as secondary current paths (or ``sneak" paths) in the absence of access selectors (NMOS) or self-rectification\cite{chen2022study}, and static power dissipation\cite{huang2024memristor}.

Hafnia-based ferroelectric resistive weights address these challenges. In these polycrystalline thin films, a permanent electrical dipole exists in each grain, which can be reversibly flipped by applying an electric field. The resistance of the device is set with electrical pulses, controlling the fraction of ferroelectric domains oriented towards a given electrode.  

Regarding the sneak path, their compatibility with CMOS allows a ``1T-1R" (1 Transistor-1 Resistor) embodiment. Three-terminal multi-bit ferroelectric field-effect transistors (FeFETs) were demonstrated in a \qty{28}{\nano\metre} HKMG process\cite{de2022read} and in the back-end-of-line (BEOL) of a \qty{22}{\nano\metre} process\cite{dutta2020monolithic}. Ferroelectric capacitors were integrated in the BEOL of a \qty{130}{\nano\metre} process \cite{9265061} and a \qty{180}{\nano\metre} process \cite{11562848}. Recently, two-terminal resistive synapses were co-integrated with a \qty{180}{\nano\metre} process\cite{begon2024back}. In this proof-of-concept, the integration of ferroelectric hafnia using HfO\textsubscript{2}-ZrO\textsubscript{2} multilayers was demonstrated in the middle-of-line above W vias, requiring dedicated foundry engineering runs. 

In this work, we integrate ferroelectric synaptic weights, consisting of W/TiN/WO\textsubscript{x}/[HfO\textsubscript{2}-ZrO\textsubscript{2}]/Mo/SiO\textsubscript{2} directly on as-delivered, commercial XFAB XP018 CMOS dies, using BEOL-compatible post-processing fabrication methods.

\section{Materials and Methods}\label{sec:matmet}

The as-delivered CMOS dies expose AlCu metallisation through openings in the \qty{2}{\micro\metre} thick SiN passivation. Post-processing steps aim at the integration of ferroelectric synaptic weights in the BEOL. First, \qty{200}{\nano\metre} of SiO\textsubscript{2} was deposited by RF sputtering, isolating the AlCu pads from the devices.

\begin{figure}[b]
    \centering
\begingroup%
\def\pfu{0.111603}
\def\pfload#1#2{\fontfamily{phv}\fontseries{#2}\fontshape{n}\fontsize{#1pt}{#1pt}\selectfont}%
\def\pfcdot{\textperiodcentered}%
\newdimen\pflen%
\def\pfsize#1#2{%
  \pfload{10}{#1}\pflen=\fontcharht\font`H\relax
  \pgfmathsetmacro\pfsz{10*#2*\pfu/\pflen}%
  \pgfmathsetmacro\pfsubsz{0.655*\pfsz}%
  \pgfmathsetlengthmacro\pfdrop{0.364*#2*\pfu}%
  \def\pfseries{#1}\pfload{\pfsz}{#1}}%
\def\pfsub#1{\raisebox{-\pfdrop}{\pfload{\pfsubsz}{\pfseries}#1}}%
\definecolor{pfline}{HTML}{042433}%
\definecolor{pfoxl}{HTML}{082248}%
\definecolor{pfdk}{HTML}{262626}%
\definecolor{pfmetal}{HTML}{D1D1D1}%
\definecolor{pfox}{HTML}{CCDDF5}%
\definecolor{pfsio}{HTML}{CAEEFB}%
\definecolor{pfbl}{HTML}{679AE2}%
\definecolor{pfild}{HTML}{96DCF8}%
\definecolor{pfsub}{HTML}{156082}%
\definecolor{pfmo}{HTML}{E6B446}%
\definecolor{pftin}{HTML}{CCCC00}%
\definecolor{pfwo}{HTML}{7030A0}%
\definecolor{pfhf}{HTML}{4AFFD1}%
\definecolor{pfzr}{HTML}{00CB98}%
\definecolor{pfgate}{HTML}{FFFF00}%
\definecolor{pfxfab}{HTML}{92D050}%
\definecolor{pfdash}{HTML}{C2DF3D}%
\definecolor{pfmoleg}{HTML}{FFC000}%
\definecolor{pfaldleg}{HTML}{6CD5B9}%
\definecolor{pfgrey}{HTML}{BFBFBF}%
\definecolor{pflabel}{HTML}{7F7F7F}%
\begin{tikzpicture}[x=\pfu pt, y=-\pfu pt,
    o/.style={draw=pfline, line width=0.558pt, line join=miter},
    lbl/.style={anchor=base west, inner sep=0pt, outer sep=0pt},
    leg/.style={lbl, font=\pfsize{m}{55}\color{black}},
    box/.style={draw=pfline, line width=0.781pt}]
\useasboundingbox (0,0) rectangle (2258,1058);

\fill[black] (32,14) rectangle (42,998);
\fill[black] (22,997.5) -- (52,997.5) -- (37,1027) -- cycle;
\foreach \k/\c in {0/pfsio, 1/pfmoleg, 2/pfaldleg, 3/pfwo, 4/pftin, 5/pfgrey,
                   6/black, 7/black, 8/black, 9/black, 10/pfgrey}
  \filldraw[box, fill=\c] (3,50+85.8*\k) rectangle (72,105+85.8*\k);
\node[leg] at (108,107.0) {SiO\pfsub{2} RF sputtering};
\node[leg] at (111,192.9) {Mo DC sputtering};
\node[leg] at (105,278.8) {ALD \qty{300}{\celsius} 5 \pfcdot{} [HfO\pfsub{2}-ZrO\pfsub{2}]};
\node[leg] at (103,364.7) {ALD \qty{360}{\celsius} WO\pfsub{x}};
\node[leg] at (109,450.6) {ALD \qty{300}{\celsius} TiN};
\node[leg] at (109,536.5) {W DC sputtering};
\node[leg] at (110,622.4) {RTA \num{450} or \qty{500}{\celsius}, \qty{60}{\second}};
\node[leg] at (109,708.3) {Top electrode definition (RIE)};
\node[leg] at (111,794.2) {Bottom electrode opening (RIE)};
\node[leg] at (110,880.1) {Passivation + VIA};
\node[leg] at (111,966.0) {Metallisation and PAD definition};

\node[lbl, font=\pfsize{m}{48}\color{pflabel}] at (1015,91) {Contact to NMOS};
\node[lbl, font=\pfsize{m}{48}\color{pflabel}] at (1674,90) {Single FE device};
\begin{scope}[pflabel, line width=0.5pt, line cap=round]
  \def\pfbrace#1#2#3{%
    \draw (#1,126) .. controls (#1,119) and (#1+4,115.5) .. (#1+14,115.5)
      -- (#2-12,115.5) .. controls (#2-4,115.5) and (#2,112) .. (#2,105)
      .. controls (#2,112) and (#2+4,115.5) .. (#2+12,115.5)
      -- (#3-14,115.5) .. controls (#3-4,115.5) and (#3,119) .. (#3,126);}
  \pfbrace{1440}{1500.5}{1562}
  \pfbrace{1634.5}{1769.5}{1904.5}
\end{scope}

\filldraw[o, draw=pfdk, fill=pfbl] (1434.5,596) rectangle (1936,646);
\filldraw[o, draw=pfdk, fill=pfmetal] (1489,596) -- (1549,596) -- (1537,646) -- (1502,646) -- cycle;
\filldraw[o, fill=pfild] (1434.5,646) rectangle (1936,749);
\filldraw[o, fill=pfsio] (1434.5,749) rectangle (1936,840);
\filldraw[o, fill=pfsub] (1434.5,840) rectangle (1936,894);
\foreach \a/\b in {1653/1726, 1803/1876}
  \filldraw[o, fill=black] (\a,840) -- (\a,858)
    arc[start angle=180, end angle=90, radius=13] -- (\b-13,871)
    arc[start angle=90, end angle=0, radius=13] -- (\b,840) -- cycle;
\filldraw[o, fill=pfgate] (1731,788) rectangle (1800,840);
\filldraw[o, fill=pfmetal] (1672,749) rectangle (1701,840);
\filldraw[o, fill=pfmetal] (1828,749) rectangle (1857,840);
\foreach \a/\b in {1486/1554, 1655/1720, 1732/1797, 1810/1875}
  \filldraw[o, fill=pfmetal] (\a,697) rectangle (\b,749);
\filldraw[o, fill=pfmetal] (1503,646) rectangle (1537,697);

\filldraw[o, draw=pfoxl, fill=pfox] (1434.5,178) rectangle (1936,596);
\foreach \a/\b in {1459.5/1490, 1550/1581, 1631/1904} {
  \filldraw[o, draw=pfoxl, fill=pfsio] (\a,521.5) rectangle (\b,596);
  \filldraw[o, fill=pfmo] (\a,456) rectangle (\b,521.5);
  \foreach \t/\u/\c in {341/360/pfhf, 360/380/pfzr, 380/399/pfhf,
                        399/418.5/pfzr, 418.5/437/pfhf, 437/456/pfzr}
    \filldraw[o, fill=\c] (\a,\t) rectangle (\b,\u);
}
\filldraw[o, fill=pfmo]  (1662,209) rectangle (1783,246);
\filldraw[o, fill=pftin] (1662,246) rectangle (1783.5,308.5);
\filldraw[o, fill=pfwo]  (1662,308.5) rectangle (1784,341);
\filldraw[o, fill=pfmetal] (1490,341) rectangle (1550,596);
\foreach \pa/\pb/\ca/\cb/\bot in {1440/1562/1500.5/1539.5/341,
                                   1679/1762/1701.5/1739.5/209,
                                   1801/1884.5/1823.5/1861.5/466}
  \filldraw[o, fill=pfmetal] (\pa,140) -- (\pb,140) -- (\pb,178) -- (\cb,178)
    -- (\cb,\bot) -- (\ca,\bot) -- (\ca,178) -- (\pa,178) -- cycle;
\node[lbl, font=\pfsize{m}{48}\color{black}] at (1669,301) {TiN};
\node[lbl, font=\pfsize{m}{48}\color{black}] at (1645,512) {Mo};

\begin{scope}[pfdash, line width=0.781pt, dash pattern=on 3.0691pt off 2.2879pt]
  \draw (1375.5,596) -- (1984.5,596);
  \draw (1378.5,997) -- (1984,997);
  \draw (1375.5,615.5) -- (1375.5,997);
  \draw (1984.5,612.5) -- (1984.5,997);
\end{scope}
\node[lbl, font=\pfsize{b}{55}\color{pfxfab}] at (1476,970) {XFAB};
\node[lbl, font=\pfsize{b}{55}\color{pfxfab}] at (1688,970) {XP018};
\end{tikzpicture}%
\endgroup%
    \caption{Process flow for BEOL integration and schematic cross-section of NMOS and single device test structures. All ALD processes were performed without breaking vacuum. Each layer was patterned using optical lithography.}
    \label{fig:Process}
\end{figure}
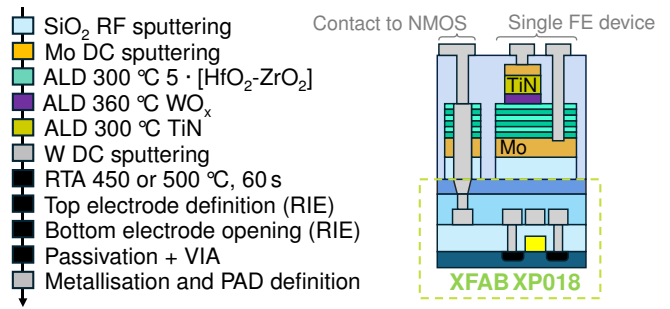

The bottom electrode was then deposited, consisting of \qty{20}{\nano\metre} of Mo deposited by DC sputtering. 
Using Plasma-Enhanced Atomic Layer Deposition (PEALD), five super-cycles were deposited at \qty{300}{\celsius}: alternating five cycles with 
\hfald{} and O\textsubscript{2}, and ten cycles with 
\zrald{} and O\textsubscript{2}, forming a [HfO\textsubscript{2}-ZrO\textsubscript{2}]\textsubscript{5} multi-layer (HZO-ML). The thickness of the HZO-ML stack is \qty{5}{\nano\metre}.

In the same deposition, 45 cycles, corresponding to \qty{2}{\nano\metre} of WO\textsubscript{x}, were deposited at \qty{360}{\celsius} without breaking vacuum.
Still without breaking vacuum, \qty{3}{\nm} of TiN was deposited by ALD, followed by \qty{20}{\nm} of DC-sputtered W, forming the top electrode.
Crystallisation was performed by rapid thermal annealing (RTA) at \qty{450}{\celsius} or \qty{500}{\celsius} for \qty{60}{\second}, at a ramp rate of \qty{30}{\celsius\per\second}, in an argon atmosphere.
The top electrode was defined by optical lithography and reactive ion etching (RIE) using a low power SF\textsubscript{6} plasma.

To connect the electrodes of the ferroelectric device to the CMOS, vias were formed using RIE and ion beam etching (IBE), with \qty{100}{\nano\metre} RF sputtered SiO\textsubscript{2} and \qty{150}{\nano\metre} DC sputtered W forming the intermetal dielectric and metallisation, respectively. Test structures allow for the direct characterisation of synaptic weights ranging from \qtyrange{2000}{5000}{\micro\metre\squared} and of the NMOS transistors of dimensions $W/L = 0.22/0.18$, $0.5/0.18$ and $2/2$ with $W$ and $L$ in \si{\micro\metre}, as represented in Fig.~\ref{fig:Process}.

In all electrical measurements the bias is applied to the top electrode using a Keysight B2912B SMU, an aiXACCT Systems TF Analyzer 3000, or a Keysight B1500A Parameter Analyzer.

\section{Results}\label{sec:res}

\subsection{BEOL compatibility}

The annealing step in the fabrication process is the bottleneck for the BEOL integration of FE devices. The thermal budget for crystallisation further increases as the film thickness decreases\cite{park_2013_anneal}. The choice of a large node (\qty{180}{\nano\metre}) is motivated here by a higher robustness to high temperature processes. 
To confirm that the NMOS transistors were not damaged during the fabrication process, particularly the RTA treatment, transfer and output characteristics were measured after pad opening for both processed (see Fig.~\ref{fig:Process}) and unprocessed reference transistors. The fully processed coupons were annealed at \qty{500}{\celsius}, conventionally exceeding the maximal thermal budget for BEOL integration but explored in this work as a process parameter.

\begin{figure}[b]
  \centering
  \includegraphics[]{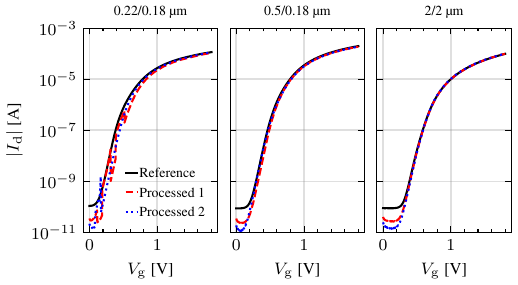}
  \caption{$I_d$-$V_g$ sweeps for the different sizes of NMOS transistors measured on fully processed chips including a \qty{500}{\celsius} annealing step, at $V_\mathrm{d}$ of \qty{1.8}{\volt}.}
  \label{fig:cmos}
\end{figure}

Id-Vg sweeps are shown for all samples in Fig.~\ref{fig:cmos}.
At $V_g=0$, the measured drain current of the processed devices is at or below the \qty{20}{\pico\ampere} measurement floor, with no increase resolved relative to the reference devices. The resulting metrics are listed in Table~\ref{tab:cmos}. The subthreshold swing (SS) is consistent across all transistors, and the \qty{2}{\micro\metre} channel devices show $<$\qty{2}{\percent} change in saturation current. The shorter channel transistors show saturation currents reduced by \qty{6}{\percent} and threshold voltages increased by up to \qty{60}{\milli\volt}. The as-delivered die-to-die variability was not measured. The transistors are fully functional after final processing of the samples, demonstrating the BEOL compatibility of the processing steps.

\begin{table}[H]
  \centering
  \caption{Comparison of NMOS transistor characteristics between an untreated reference die and two fully processed dies, including \qty{500}{\celsius} thermal annealing.}
  \label{tab:cmos}
  \setlength{\tabcolsep}{3.5pt}
  \begin{tabular}{@{}ll
                  S[table-format=3.0]
                  S[table-format=2.0]
                  S[table-format=1.1]
                  S[table-format=3.1]@{}}
    \toprule
    $W/L$ [\si{\micro\metre}] & Die & {$V_\mathrm{th}$ [\si{\milli\volt}]} & {SS [\si{\milli\volt/dec}]}
      & {$R_\mathrm{on}$ [\si{\kilo\ohm}]} & {$I_\mathrm{d,sat}$ [\si{\micro\ampere}]} \\
    \midrule
    \multirow{3}{*}{0.22/0.18}
      & Ref.  & 394   & 83  & 3.5  & 115.2 \\
      & P1    & 421   & 89  & 3.7  & 107.9 \\
      & P2    & 453   & 86  & 3.6  & 112.5 \\
    \midrule
    \multirow{3}{*}{0.5/0.18}
      & Ref.  & 507   & 89  & 2.1  & 198.0 \\
      & P1    & 531   & 86  & 2.2  & 187.9 \\
      & P2    & 516   & 92  & 2.2  & 188.6 \\
    \midrule
    \multirow{3}{*}{2/2}
      & Ref.  & 584   & 97  & 4.9  & 98.4 \\
      & P1    & 588   & 84  & 4.9  & 99.0 \\
      & P2    & 588   & 83  & 4.8  & 100.3\\
    \bottomrule
  \end{tabular}
\end{table}

\begin{figure}[h]
  \centering
  \includegraphics[]{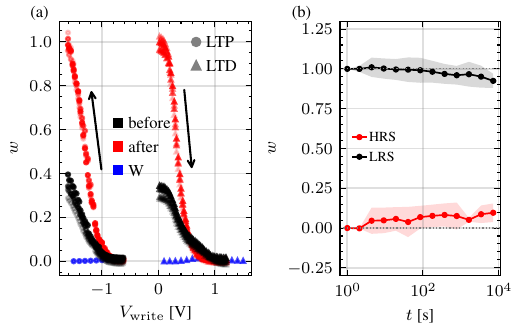}
  \caption{Weight update and retention for \qty{450}{\celsius} devices. (a) Comparison of the weight update of one device before and after wake-up, over four cycles. The cycle number is encoded in the opacity, with the first cycle being the faintest. The conductance is normalised to the range after wake-up. The weight update for a W device annealed at \qty{450}{\celsius} is displayed in blue. (b) Conductance retention for LRS (HRS) after a \qty{-1.5}{\volt} (\qty{1.2}{\volt}) write pulse, for 3 devices. Both states drift towards each other by \qty{7.5 \pm 0.4}{\percent} (\qty{9.5 \pm 0.6}{\percent}) within \qty{2}{\hour}.}
  \label{fig:wakeup}
  \label{fig:retention}
\end{figure}

\begin{figure}[h]
  \centering
  \includegraphics[]{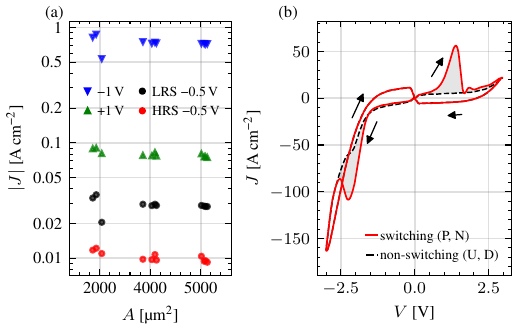}
  \caption{Evidence for homogeneous, ferroelectric switching in \qty{450}{\celsius} devices. (a) Non-switching current density of all devices versus the device areas. The current density and dynamic range are independent of area and the rectification ratio is \num{9}. The horizontal spread for each area is for visibility only. (b) PUND measurement at \qty{100}{\kilo\hertz}, averaged over 8 cycles on one device, resulting in $2P_\mathrm{r}$ of \qty{22.1}{\micro\coulomb\per\centi\metre\squared}. The shaded area represents the switched charge.}
  \label{fig:area}
  \label{fig:pund}
\end{figure}

\subsection{Electrode choice and synaptic functionality}


W and TiN are known to promote the crystallisation of hafnia in the ferroelectric phase\cite{Wang_1,10.1063/5.0029516,10.1021/acsami.4c15934}, providing mechanical stress during the annealing. However, in contrast to previous work on partially processed wafers, devices based on W and TiN electrodes (as in\cite{begon2024back}) fail to exhibit synaptic functionality on fully processed wafers (blue symbols in Fig.~\ref{fig:wakeup}(a)). To address this problem, molybdenum, a promising electrode candidate for hafnia \cite{10543058,8731272}, is investigated. 

At the top electrode, the role of the WO\textsubscript{x} interlayer is twofold. First, it supplies oxygen during the crystallisation\cite{YANG2023107565, 10.1021/acsami.4c18056}. Second, it contributes to the synaptic weight functionality \cite{begon2022effect}. 

Hafnia ferroelectric memories often require a ``wake-up" procedure \cite{10.1021/acsaelm.0c00671}: it refers to the need to cycle devices under an electric field before they become operational. Weight update experiments are performed before and after wake-up: after programming the device with a pulse of amplitude $V_{\mathrm{write}}$ and duration \qty{500}{\micro\second}, the conductance is measured non-destructively at \qty{-0.5}{\volt}. 
Fig.~\ref{fig:wakeup}(a) shows four weight update experiments on a pristine device (black symbols) and on the same device after 1000 cycles (red symbols) from \qtyrange{-1.6}{1.2}{\volt}. 
The devices can be successfully programmed to a high and a low resistive state (HRS, LRS), as well as intermediate states, emulating synaptic functionality: long-term potentiation (LTP) and long-term depression (LTD). In the following, we define the normalised conductance, or weight, as $w = G_\mathrm{norm} = (G - G_\mathrm{HRS})/(G_\mathrm{LRS} - G_\mathrm{HRS})$.
In the first four experiments, the device already exhibits LTP and LTD. After 1000 cycles, the dynamic range increases by a factor of \num{3}, and the median cycle-to-cycle variability falls from \qty{5}{\percent} to \qty{1.5}{\percent} of the window. 
Finally, short-term relaxation effects are investigated. Immediately after a \qty{-1.5}{\volt} (\qty{1.2}{\volt}) programming pulse, the read conductance is repeatedly measured at \qty{-0.5}{\volt} over two hours. 
The LRS (HRS) shows a relaxation drift of \qty{7.5 \pm 0.4}{\percent} (\qty{9.5 \pm 0.6}{\percent}), see Fig.~\ref{fig:retention}(b), which could originate from ionic contributions at the Mo or WO\textsubscript{x} interfaces.

\subsection{Weight update mechanisms}

To verify that the weight update mechanism originates from ferroelectricity, three experiments are conducted.

First, DC I-V measurements are performed on devices of various areas. Fig.~\ref{fig:area}(a) shows that the current density is constant. This indicates that the change in conductance occurs homogeneously, in contrast to RRAM based on filament formation. For the \qty{5000}{\micro\metre\squared} area devices, the device-to-device variability is \qty{1.8}{\percent}. The rectifying ratio is 
9 at $\pm$\qty{1}{\volt}.

Second, a PUND (Positive Up Negative Down) test\cite{pund} is performed on a BEOL integrated device. The current density is represented in Fig.~\ref{fig:pund}(b). During the first pulse of each polarity (red lines), a switching peak is observed, but not during the second pulse (black lines). The  difference between these two responses provides the charge that flowed from one electrode to the other during polarisation reversal per unit area, which is the ferroelectric remanent polarisation\cite{pund}. 

\begin{figure}[h]
  \centering
  \includegraphics[]{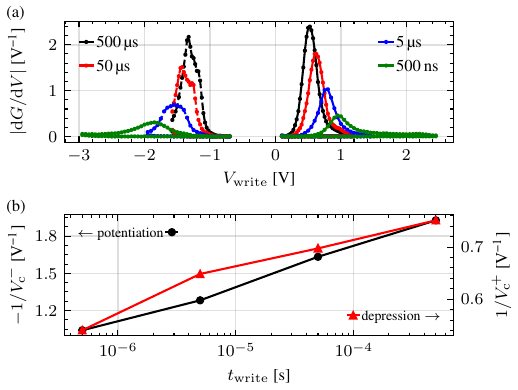}
  \caption{LTP-LTD with different writing pulse widths. (a) Conductance change per voltage step for write voltage pulse widths between \qty{500}{\nano\second} and \qty{500}{\micro\second}. (b) Inflection point $V_c$ of the switching curve (giving $\max\left|\mathrm{d}G_\mathrm{norm}/\mathrm{d}V\right|$) as a function of the pulse width $t_{\mathrm{write}}$.}
  \label{fig:tfa}
\end{figure}

Third, the weight update is evaluated under different programming durations. The derivative of the conductance as a function of the programming amplitude is represented in Fig.~\ref{fig:tfa}(a). It shows a maximum at $|V_{\mathrm{c}}|$. As shown in Fig.~\ref{fig:tfa}(b), $1/|V_{\mathrm{c}}|$ increases linearly as $\log(t_{\mathrm{write}})$ increases. 
This observation is consistent with the Merz law\cite{merz} describing domain reversal in ferroelectric thin films.

Together, these experiments support that the weight update originates from ferroelectric domain reversal in BEOL integrated W/TiN/WO\textsubscript{x}/HZO-ML/Mo/SiO\textsubscript{2}/XP018 synaptic weights, likely paired with charge trapping and ionic redistribution\cite{begon2022effect}.

\subsection{Effect of the weight update characteristics on system level performance}

\begin{figure}[b]
  \centering
  \input{Figures/pgf/ltpltd_full.pgf}
  \includegraphics[]{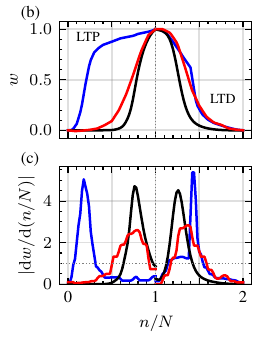}
  \caption{Weight update of the ferroelectric synapses over amplitude ramped pulses, averaged over five cycles per device.
    (a) Normalised conductance $w$ as a function of write voltage.
    (b) Normalised conductance as a function of normalised pulse number $n/N$, for potentiation (0-1) and depression (1-2). Each branch is scaled to its own first and last pulse.
    (c) Derivative of the data in (b), the normalised conductance change per normalised pulse. The \qty{450}{\celsius} devices and the reference~\cite{begon2024back} reach similar peak slopes of 4.5 to 5.4, the \qty{500}{\celsius} devices about half of that.}
  \label{fig:sym}
\end{figure}

Finally, the linearity of the LTP and LTD is quantified for an amplitude-incremented pulse scheme. 
Write pulses from \qtyrange{-1.6}{1.2}{\volt}, with \qty{25}{\milli\volt} step size and a \qty{500}{\micro\second} write pulse width, are applied to the devices. Four cycles are repeated consecutively, showing cycle-to-cycle variability below \qty{1}{\percent}. 
The switching loops of the \qty{450}{\celsius} devices exhibit an offset of \qty{-0.4}{\volt} between the centre-points of the two branches. This offset is independent of write pulse width, see Fig.~\ref{fig:tfa}, indicating a field related origin rather than one grounded in switching kinetics.  
One possible cause is the difference in the work functions of the electrodes. The switching asymmetry may reflect differences in electrode screening, interfacial charge or the ferroelectric phase distribution. To investigate the influence of annealing temperature, a second chip with the same nominal stack was annealed at \qty{500}{\celsius}. The resulting reduction in loop offset and more gradual weight updates indicate a change in switching behaviour, although its microscopic origin is not established by the present measurements. This reduces the offset from \qty{-0.4}{\volt} to \qty{-0.1}{\volt} and narrows the write window from \qty{2.8}{\volt} to \qty{2.4}{\volt}: the negative limit increases from \qty{-1.6}{\volt} to \qty{-1.2}{\volt}, while the positive remains at \qty{1.2}{\volt}. Weight update experiments reveal that the cycle-to-cycle variation is \qty{2.3}{\percent} over \num{4} cycles. The \qty{500}{\celsius} devices show more gradual weight updates, with a peak slope
below \num{3} against \numrange{4.5}{5.4} for the \qty{450}{\celsius} devices
and the reference~\cite{begon2024back} (Fig.~\ref{fig:sym}). 

Using the PyTorch toolkit ``IBM Analog Hardware Acceleration Kit"\cite{9458494}, one can compare simulated system level performance of the synaptic weights. 
The normalised LTP and LTD are fitted with a piecewise function (\texttt{PiecewiseStepDevice} model). The training  of a neural network comprising two hidden layers (in total 784, 256, 128 and 10 nodes) for the MNIST classification task is simulated. The batch size is 64 and the learning rate is 0.5 through 50 epochs.
Table~\ref{tab:aihwtk} shows the final accuracy for the W/TiN/WO\textsubscript{x}/HZO-ML/Mo/SiO\textsubscript{2}/XP018 chips annealed at \qty{450}{\celsius} and \qty{500}{\celsius}. The listed parameters used for the simulations are defined in the toolkit documentation \cite{9458494}: for example, \texttt{dw\_min\_std} and \texttt{dw\_min\_dtod} are functions of the cycle-to-cycle and device-to-device variability, respectively, and of the minimal weight update. The parameters are taken equal for both chips, such that only the effect of the update-curve shape is evaluated. 
The mixed-precision stochastic gradient descent (MP-SGD) scheme emulates hardware comprising digital cores, in which the weight update is accumulated over epochs until it reaches a certain threshold. This scheme is compatible with the amplitude-incremented pulse scheme. It allows the neural network to exhibit learning accuracy of \qty{97}{\percent} for both annealing temperatures. In contrast, the plain SGD scheme reflects the case of an analog array with state-aware drivers in hardware, in which the weights are updated at each training example. In the case of ideal devices, the chip annealed at \qty{500}{\celsius} reaches \qty{87}{\percent} accuracy, against \qty{80}{\percent} for the chip annealed at \qty{450}{\celsius}. The gap is even larger when simulations account for the real cycle-to-cycle and device-to-device variations, with \qty{82}{\percent} and \qty{69}{\percent}. These results highlight how subtle changes in processing conditions can significantly impact the device behaviour and consequently the system level performance.

\newcolumntype{C}[1]{>{\centering\arraybackslash}m{\dimexpr#1-2\tabcolsep\relax}}
\newcommand{\thick}{0.5pt}
\begin{table}[t]
\centering
    \caption{Neural network simulations: parameters and results}
    \label{tab:aihwtk}
    \setlength{\tabcolsep}{1.5pt}      
    \renewcommand{\arraystretch}{1.66} 
\begin{tabular}{C{70pt}C{27pt}C{29pt}C{42pt}C{31pt}C{46pt}}
    \toprule
    \textbf{Parameter} & \textbf{ideal} & \textbf{real} & \textbf{Training} & \textbf{RTA} & \textbf{Accuracy} \\
    \midrule
    \texttt{dw\_min\_std}    & 0 & 0.041 & \multirow{2}{=}{\centering MP-SGD, real} & \qty{450}{\celsius} & \qty{97 \pm 0.1}{\percent} \\
    \cline{5-6}
    \texttt{dw\_min\_dtod}   & 0 & 0.494 &                                          & \qty{500}{\celsius} & \qty{97 \pm 0.1}{\percent} \\
    \Xcline{4-6}{\thick}
    \texttt{w\_min\_dtod}    & 0 & 0.113 & \multirow{2}{=}{\centering SGD, ideal}   & \qty{450}{\celsius} & \qty{80 \pm 1.7}{\percent} \\
    \cline{5-6}
    \texttt{w\_max\_dtod}    & 0 & 0.115 &                                          & \qty{500}{\celsius} & \qty{87 \pm 0.2}{\percent} \\
    \Xcline{4-6}{\thick}
    \texttt{up\_down\_dtod}  & 0 & 0.075 & \multirow{2}{=}{\centering SGD, real}    & \qty{450}{\celsius} & \textbf{\qty{69 \pm 1.7}{\percent}} \\
    \cline{5-6}
    number of seeds & 3 & 3     &                                          & \qty{500}{\celsius} & \textbf{\qty{82 \pm 2.9}{\percent}} \\
    \bottomrule
\end{tabular}

\end{table}

\section{Summary and Outlook}

In summary, we integrated ferroelectric synaptic weights based on a
W/TiN/WO\textsubscript{x}/HZO-ML/Mo stack directly on as-delivered, commercial
XFAB XP018 CMOS dies.
The NMOS transistors maintain a stable subthreshold swing
and saturation current, varying by at most \qty{6}{\percent}, even in the worst-case \qty{500}{\celsius} annealing. Ferroelectric switching was established by three independent observations: an area-independent
current density, a PUND response with $2P_r$ of \qty{22.1}{\micro\coulomb\per\centi\metre\squared}, and switching kinetics following the Merz law. The devices show analog
potentiation and depression with cycle-to-cycle variability below
\qty{1.5}{\percent} and device-to-device variability of \qty{1.8}{\percent}. Raising the
annealing temperature from \num{450} to \qty{500}{\celsius} improves the
LTP/LTD linearity and allows \qty{1.2}{\volt} operation. In simulated MNIST training, this
raises the accuracy of an analog SGD implementation with state-aware drivers by 13 percentage points. Small
changes in processing conditions therefore propagate directly to
system-level performance. Future work will focus on characterising 1T-1R cells. Further improvement of the update linearity, symmetry, and retention, for instance through electrode work-function engineering, is required to close the gap between analog SGD and mixed-precision training. 

\section*{Acknowledgments}

The authors acknowledge Prof. Erika Covi and Bojian Zhang for designing and providing the CMOS wafer, and the Binnig and Rohrer Nanotechnology Center (BRNC). Research funded through the SNSF Starting Grant ROSUBIO (218438), by the Swiss State Secretariat for Education, Research, and Innovation (SERI) through the SwissChips research project and by Horizon Europe Chips JU through ViTFOX (101194368).

\bibliographystyle{IEEEtran}
\bibliography{iscas_2027_v2_bib.bib}

\end{document}